\documentclass[prl,twocolumn,superscriptaddress]{revtex4-2}
\usepackage{amsfonts}
\usepackage{amssymb}
\usepackage{amsmath}
\usepackage{graphicx}
\usepackage{dcolumn}
\usepackage{appendix}
\usepackage{bm}
\usepackage{units}
\usepackage{txfonts}
\usepackage{multirow}
\usepackage{CJKutf8}
\usepackage{subfigure}
\usepackage{color}
\usepackage{soul}
\usepackage{ulem}
\usepackage{url}
\usepackage[colorlinks,linkcolor=blue,anchorcolor=blue,citecolor=blue,urlcolor=blue]{hyperref}
\usepackage{array}
\begin{document}

\title{Skyrmion pinning by disks} 

\author{X. Gong}
\affiliation{Physics Department, The Hong Kong University of Science and 
Technology, Clear Water Bay, Kowloon, Hong Kong}

\author{K. Y. Jing}
\affiliation{Physics Department, The Hong Kong University of Science and 
Technology, Clear Water Bay, Kowloon, Hong Kong}

\author{J. Lu}
\affiliation{College of Physics Science and Technology, Yangzhou University, 
Yangzhou 225002, People's Republic of China}

\author{X. R. Wang}
\email[Corresponding author: ]{phxwan@ust.hk}
\affiliation{Physics Department, The Hong Kong University of Science and 
Technology, Clear Water Bay, Kowloon, Hong Kong}
\affiliation{HKUST Shenzhen Research Institute, Shenzhen 518057, China}
\date{\today}

\begin{abstract}
High precision skyrmion pinning by an intentionally created magnetic structure 
is important in skyrmion manipulation. Here, we consider skyrmion pinning by 
various types of disks. Among other findings, we clarify that, in terms of 
pinning position and pinning potential landscape, one needs to distinguish 
thin-wall skyrmions from thick-wall skyrmions because of fundamental 
differences. A skyrmion wall prefers areas with a weaker exchange stiffness, or 
a larger DMI constant, or a smaller magnetic anisotropy while a skyrmion core 
experiences only the magnetic anisotropy, not exchange stiffness and DMI. 
Depending on disk types, skyrmion types, and the relative size of disk 
(to skyrmion), a skyrmion can be pinned at the symmetric (center) or 
asymmetric (off-center) point of a disk. In case that a skyrmion 
is pinned by a local energy minimum, thermal agitation can depin a skyrmion, 
and the pinning lifetime follows an Arrhenius law. Interestingly, when the disk 
size is comparable to the skyrmion size, the skyrmion deforms greatly 
such that the skyrmion will shrink or expand to fill the whole disk. 
These findings should be important for skyrmion manipulations. 
\end{abstract}
\maketitle

\section{introduction}
Magnetic skyrmions, topologically non-trivial spin textures, have been observed 
in chiral magnets with Dzyaloshinskii–Moriya interactions (DMIs) \cite{YuXZ2010,
Muhlbauer2009,Bogdanov2006,JiangW2015,Sampaio2013,Heinze2011,Boulle2016}. 
Their non-trivial topology, small size and ultralow driving force make them promising 
information carriers \cite{Fert2013,Tomasello2014,Rohart2013,Nagaosa2013,Parkin2020,
Klaui2016,ZhangXC2015,Martinez2018,Ding2015,Queralt2019,Iwasaki2013,Krause2016,
Kim2017,David2017}. Similar to the importance of magnetic domain wall (DW) pinning 
\cite{yuan1} in DW-based devices, skyrmion pinning by disorders and 
intentionally created nano-structures is important in skyrmionics \cite
{Dusan2017,Hanneken2016,Liu2013,Lin2013,Muller2015,Toscano2019,Song2021,
Pathak2021,Fernandes2019,depin3,Arjana2020,GongX2020,JingKY2021,Suess2020,CR2015}. 
Skyrmion pinning in nano-structures with different exchange stiffness, or 
DMI strength, or magnetic anisotropy, or saturation magnetization has been  
investigated  \cite{Fert2013,Dusan2017,Song2021,Toscano2019,Liu2013}. 
Skyrmion pinning by atomistic spin vacancies or geometrical constrains 
like film thickness modulations and semicircular notches at lateral 
boundaries has also been studied \cite{Dusan2017,Pathak2021,Suess2020}.   
It was found \cite{Dusan2017} that pinning by a small structure can occur at 
or away from the center of a nano-structure. It was also known experimentally 
and numerically that a nano-structure modulated by the exchange stiffness 
or DMI strength pins a skyrmion differently in terms of pinning position.  
A magnetic field can also be used to manipulate the pinning position 
\cite{Dusan2017, Hanneken2016}. The interaction between a nano-structure and 
a skyrmion was further classified into pinning problems and scattering problems, 
depending on whether structure size is small or larger than skyrmion size 
\cite{Toscano2019}. It was also observed that skyrmion structure may change  
before and after pinning \cite{Dusan2017,Song2021}. Despite of all of these 
accumulated knowledge, there is still a lack of a clear understanding of 
various pinning phenomena. For example, there is no explanation why a similar 
structure pins a skyrmion sometimes at the center and sometimes off-center. 
There is no explanation why a skyrmion sometimes undergoes a great deformation. 
Most of our knowledge is purely phenomenological and afterwards statement, 
lack of a simple physical picture for various skyrmion pinning configurations. 
In other words, our existing knowledge does not have a predictive/designing power. 

In this study, we systematically investigate skyrmion pinning by a disk with 
different material parameters from its embedded homogeneous chiral magnetic film.  
We find that pinning position is highly sensitive to whether the skyrmion 
has a thin-wall or a thick-wall, which can be characterized by the ratio 
of skyrmion radius $R$ to the skyrmion wall width $w$ \cite{skyrmionsize}.
This finding resolves the puzzle why some groups observed skyrmion pinning at 
center while other groups observed pinning off-center with similar structures. 
It is also found that skyrmion deformation in a small disk is negligible 
so that rigid skyrmion approximation is applicable. A simple theory is capable 
of accurately predicting the pinning position and skyrmion-disk interaction.
In the case when skyrmion is pined at a local pinning site, it can be 
depinned at finite temperature and the pinning lifetime follows an Arrhenius law.
The gigantic skyrmion structure deformation happens when the skyrmion size 
is comparable to the disk size such that deformation energy is smaller 
than the energy gain from putting whole skyrmion inside a low energy land. 

The paper is organized as follows. The model is introduced in the next section. 
Section \ref{sec3} presents our main findings, including the skyrmion-structure 
dependences of pinning position, skyrmion pinning lifetime at finite temperature,
skyrmion deformation and the pinning position as a function of disk types and 
disk size for both thin-wall and thick-wall skyrmions. The discussions 
and conclusion are given in Sec. \ref{sec4} followed by the acknowledgement.

\section{Model and Methods}
\label{sec2}
We consider a chiral magnetic thin film of thickness $d$ in $xy$-plane with a disk 
of radius $R_d$ entered at origin. The energy of ferromagnetic state of $m_z=1$ 
is used as the reference point ($E=0$), the magnetic energy of a magnetization 
structure $\vec{m}(x,y)$ (unit vector of magnetization) with an interfacial DMI is 
\begin{equation}
E=d\iint \left(\varepsilon_{\rm ex} + \varepsilon_{\rm DM}
+ \varepsilon_{\rm an} + \varepsilon_{\rm d} + \varepsilon_{\rm Ze}\right)dxdy,
\label{energy}
\end{equation}
in which $\varepsilon_{\rm ex}= A(x,y)|{\nabla} \vec{m}|^2$,
$\varepsilon_{\rm DM}= D(x,y)[ m_z\nabla\cdot\vec{m}- (\vec{m}\cdot\nabla)m_z]$,
$\varepsilon_{\rm an}= K_{\rm u}(x,y)(1-m_z^2)$,
$\varepsilon_{\rm d}= -\mu_0 M_{\rm s}\vec H_{\rm d}\cdot\vec{m}$,
and $\varepsilon_{\rm Ze}= B M_{\rm s}(1-m_z)$ are exchange, DMI, anisotropy, 
magnetic static, and the Zeeman energy densities, respectively.
Here $A(x,y)$, $D(x,y)$, $K_{\rm u}(x,y)$, $B$, $M_{\rm s}$, $\vec{H}_{\rm d}$ 
and $\mu_0$ are exchange stiffness, DMI strength, magneto-crystalline anisotropy,
perpendicular magnetic field, saturation magnetization, demagnetizing field and 
the vacuum permeability, respectively. 
We consider three types of disks, called A-, D-, and K-disks in which only $A$, 
$D$, or $K_{\rm u}$ is different inside and outside the disk, respectively. 
A disk with a weaker or a stronger exchange stiffness is called an $\text{A}_-$ or $\text{A}_+$ 
disk while a disk with a smaller or a larger DMI is called a $\text{D}_-$ or $\text{D}_+$ disk, and 
a disk with a smaller or a larger magnetic anisotropy is called a $\text{K}_-$ or $\text{K}_+$ disk. 
The material parameters of these disks are listed in the table \ref{table1}. 
For an ultra-thin film, demagnetization effect can theoretically be included in the 
effective anisotropy $K=K_{\rm u}-\mu_0M_{\rm s}^2/2$. This is a good approximation 
when the film thickness $d$ is much smaller than the exchange length \cite{skyrmionsize}. 
With that being said, the exact demagnetization field is automatically included in 
all of our MuMax3 simulations \cite{mumax3}. It is known that isolated circular 
skyrmions are metastable state when $\kappa=\pi^2 D^2/(16AK) <1$ \cite{skyrmionsize,haitao2}. 
\begin{table}[h]
\begin{tabular}{cccc}
\hline\hline\noalign{\smallskip}
Disk type & $A$($\mathrm{pJ}/\mathrm{m}$) & $D$ ($\mathrm{mJ}/\mathrm{m}^2$) 
& $K_{\rm u} (\mathrm{MJ}/\mathrm{m}^3)$ \\ \noalign{\smallskip}\hline\noalign{\smallskip}
$\text{A}_-$   &     $14.3\sim 14.7$      &  $3.58  $   &  $0.80  $          \\ 
$\text{A}_+$   &     $15.2 \sim 15.9$      &  $3.58  $   &  $0.80  $          \\ 
$\text{D}_-$   &     $15.2 $      &  $3.52\sim 3.56$   &  $0.80  $          \\ 
$\text{D}_+$   &     $15.2 $      &  $3.62\sim 3.68 $   &  $0.80  $          \\ 
$\text{K}_-$   &     $15.2 $      &  $3.58 $    &  $0.76\sim 0.78 $          \\ 
$\text{K}_+$   &     $15.2 $      &  $3.58 $    &  $0.82\sim 0.84 $          \\ 
\noalign{\smallskip}\hline
\end{tabular}
\caption{Material parameters for the six types of disks. }
\label{table1}
\end{table}

Magnetization dynamics in a magnetic field is governed by the Landau-Lifshitz-Gilbert (LLG) equation,
\begin{equation}
\frac{\partial \vec m}{\partial t} =-\gamma\vec m \times \vec H_{\rm eff} +
\alpha \vec m \times \frac{\partial \vec m}{\partial t}, 
\label{llg}
\end{equation}
where $\gamma$ and $\alpha$ are respectively gyromagnetic ratio and Gilbert 
damping constant. $\vec H_{\rm eff}=\frac{2A}{\mu_0M_{\rm s}} \nabla^2\vec m+
\frac{2K_{\rm u}}{\mu_0M_{\rm s}}m_z\hat z+H\hat z+\vec H_{\rm d}+\vec H_{\rm DM}+
\vec{h}$ is the effective field including the exchange field, the anisotropy 
field, the external magnetic field along $\hat z$, the demagnetizing field, the 
DMI field $\vec H_{\rm DM}$, and a temperature-induced random magnetic field 
 $\vec{h}=\vec{\eta}\sqrt{2\alpha k_{\rm B} T/(M_{\rm s} \mu_0 \gamma\Delta V\Delta t)}$, 
where $\Delta V$, $\Delta t$, $T$ and $\vec{\eta}$ are the cell volume, time step, the 
temperature and a random vector from a standard normal distribution whose value is 
changed after every time step, respectively \cite{mumax3,Brown}. The LLG equation describes a 
dissipative system whose energy can only decrease \cite{xrw2} in the absence 
of energy sources. 

In the polar coordinates $(r,\phi)$, the normalized magnetization is
$\vec{m}=\left(\sin \Theta \cos \Phi, \sin \Theta \sin \Phi, \cos \Theta  \right)$, 
where the polar angle $\Theta$ of the magnetization of a skyrmion centered at 
$\vec{r}_c$ can be well described by \cite{skyrmionsize,haitao2},
\begin{equation}
\Theta \left(\vec{r}\right)= 2 \arctan \left[\frac{\sinh (R/w)}{\sinh 
\left(|\vec{r}-\vec{r}_c|/w\right)}\right],
\label{profile}
\end{equation}
where $R$ is the skyrmion radius and $w$ is the skyrmion wall width. The azimuthal 
angle is $\Phi=\mathcal{V}\phi+\mathcal{H}$, where $\mathcal{V}$ is the vorticity and 
$\mathcal{H}$ is the helicity \cite{skyrmionsize,Nagaosa2013}. For a N\'{e}el-type 
skyrmion considered here, we have $\mathcal{V}=1$ and $\mathcal{H}=0$, thus $\Phi=\phi$.
The ratio $\eta = R/w$ is an important quantity characterizing skyrmion structure.  
Large $\eta$ means a relatively thin wall and a large core in which all spins align 
along the same direction, while $\eta$ around 1 describes a skyrmion of a relatively 
thick wall with a negligible core.

\begin{figure}
\centering
\includegraphics[width=8.5cm]{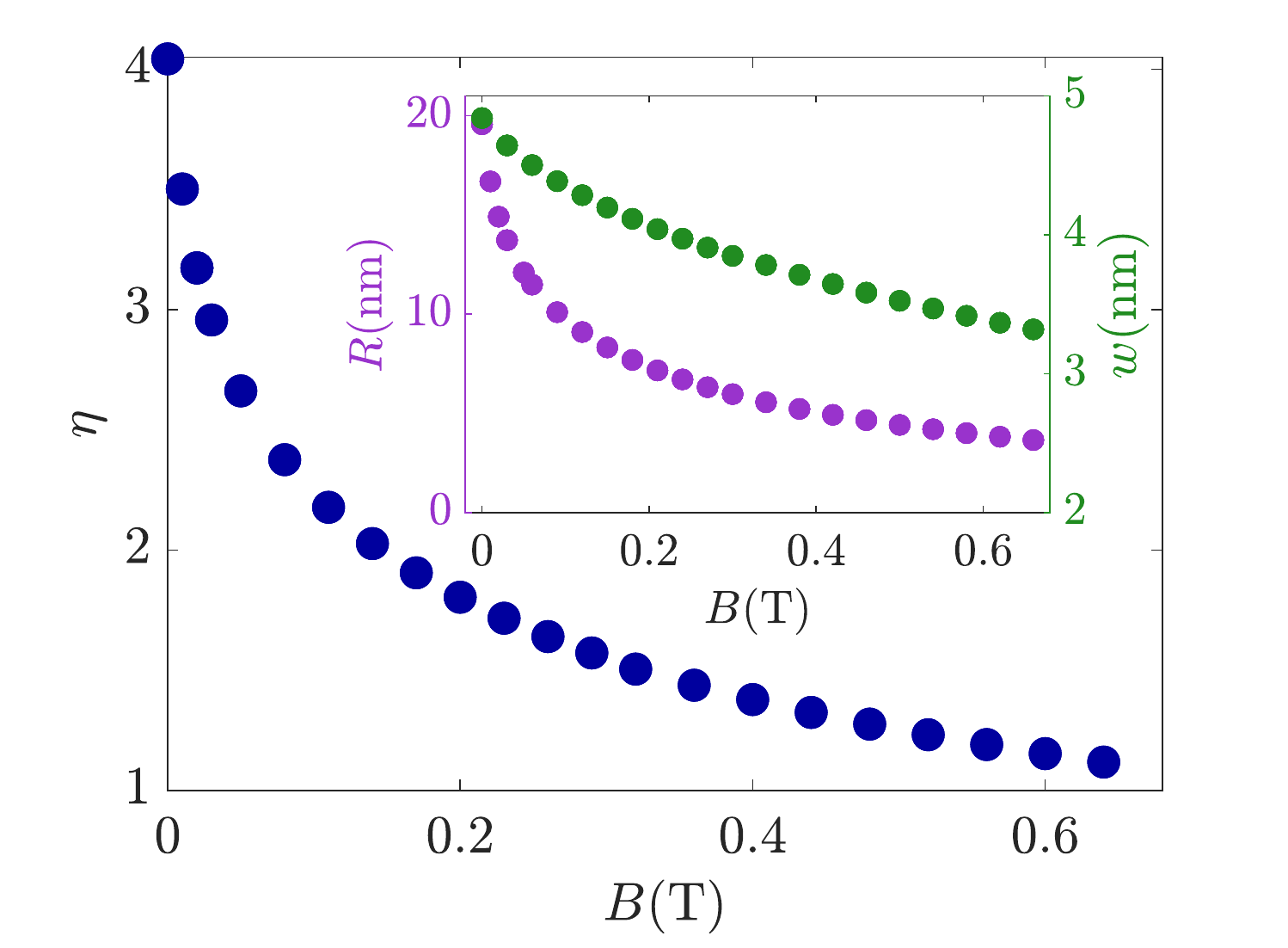}\\
\caption{Magnetic field dependence of skyrmion structure parameter $\eta$. Inset: 
field dependence of skyrmion radius $R$ and wall width $w$.}
\label{fig1}
\end{figure}

To mimic a Co/Pt thin film \cite{Sampaio2013,ptco1}, we set the parameters of films to be 
$A=15.2\, \mathrm{pJ}/\mathrm{m}$, $D=3.58 \mathrm{\,mJ}/\mathrm{m}^2 $, $K_{\rm u}=0.8 
\,\mathrm{MJ}/\mathrm{m}^3$ and $M_s=0.58 \, \mathrm{MA}/\mathrm{m}$, respectively. 
A skyrmion, whose central spin is along the $-\hat z$ direction, is created near or away 
from the disk center in a film of size $200{\rm nm} \times 200 {\rm nm}\times 1{\rm nm}$. 
$\eta$ changes with material parameters and can be controlled by an external magnetic field 
of $B \in [0,0.66] \mathrm{\,T}$ in the $+\hat z$ direction, as shown in Fig.\ref{fig1}. 
Spins prefer to align along the magnetic field $B$ (the $+z$ direction) such that the 
skyrmion core shrinks and $\eta$, as well as $R$ and $w$, decreases with the increase of 
$B$ as shown in Fig. \ref{fig1}. $R$ and $w$ can be obtained from fitting the 
magnetization profile of a stable skyrmion to Eq. \eqref{profile}. 
In our numerical studies below, the static problem such as 
pinning position, structures, and pinning energy are obtained by energy 
minimization from solver relax() in MuMax3 \cite{mumax3}. 
The full MuMax3 solver will be used to solve the LLG equation with $\alpha =1$ 
only when the stochastic process or a true pinning trajectories are wanted. 

\section{Results}
\label{sec3}
\subsection{Two types of skyrmions}

\begin{figure}
	\centering
	\includegraphics[width=8.5cm]{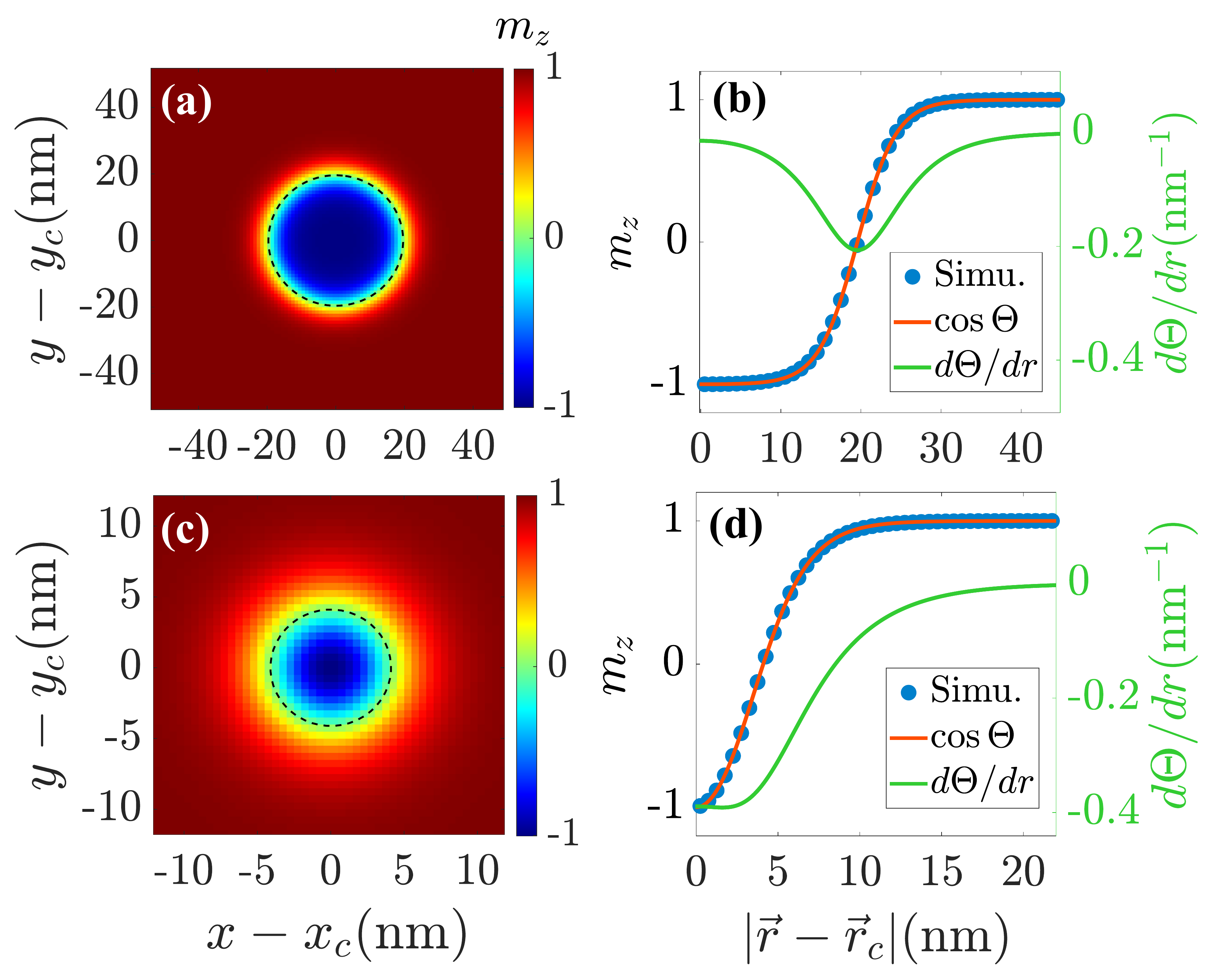}\\
\caption{Density plot (a, c) and radial distribution (b, d) of $m_z$ for a thin-wall 
skyrmion in the absence of a magnetic field with $R=19.57 \ {\rm{nm}}$, $w=4.84\ \rm{nm}$ 
(a, b) and for a thick-wall skyrmion under a perpendicular magnetic field of $B=0.56$ T 
with $R=4.10\ {\rm{nm}}$, $w=3.44\ \rm{nm}$ (c, d). Magnetization in the dark blue (the 
dark red) regions in (a) and (c) point to the $-\hat{z}$ ($+\hat{z}$), while the black 
dashed circles indicate $m_z=0$ contours. The blue dots and the red curves in (b) and 
(d) are respectively the simulation data and analytical calculation based on Eq. 
(\ref{profile}). The green curves are $d\Theta/dr$.}
	\label{fig2}
\end{figure}

$\eta$ values of Skyrmions can be tuned by a magnetic field along the $z$-direction. 
For the parameters used in this study and in the absence of a perpendicular magnetic field, 
skyrmions have a ``thin-wall" with $\eta\approx 4 \gg 1$ and a large core domain [the 
dark blue part in Fig. \ref{fig2} (a)]. The polar angle $\Theta(\vec {r})$ of the magnetization 
of a skyrmion centered at ${\vec{r}}_c=(x_c,y_c)$ is shown in Fig. \ref{fig2} (b).
The blue dots and the red curve are numerical data and Eq. (\ref{profile}) with $R=19.57 \ 
{\rm{nm}}$, $w=4.84\ \rm{nm}$, respectively. The central core of the skyrmion shrinks 
considerably under $B=0.56$T, leading to $\eta\approx 1$ as shown in Fig. \ref{fig2} (c). 
The whole skyrmion is made of a wall whose magnetization varies from point to point  
as shown by the green curve in Fig. \ref{fig2} (d) in contrast to very different 
behavior of a thin-wall skyrmion as shown by the green curve in Fig. \ref{fig2} (b). 
This is why we call skyrmions with $\eta\approx 1$ ``thick-wall". The perfect overlap of the 
numerical data with the analytical formula demonstrates the excellence of Eq. (\ref{profile}) 
in describing the magnetization profiles of both thin-wall and thick-wall skyrmions. 
Subsections below show distinct pinning behaviors for the thin-wall and thick-wall skyrmions. 

\subsection{Pinning of thin-wall skyrmions and thick-wall skyrmions}
\begin{figure*}
	\centering
	\includegraphics[width=17.5cm]{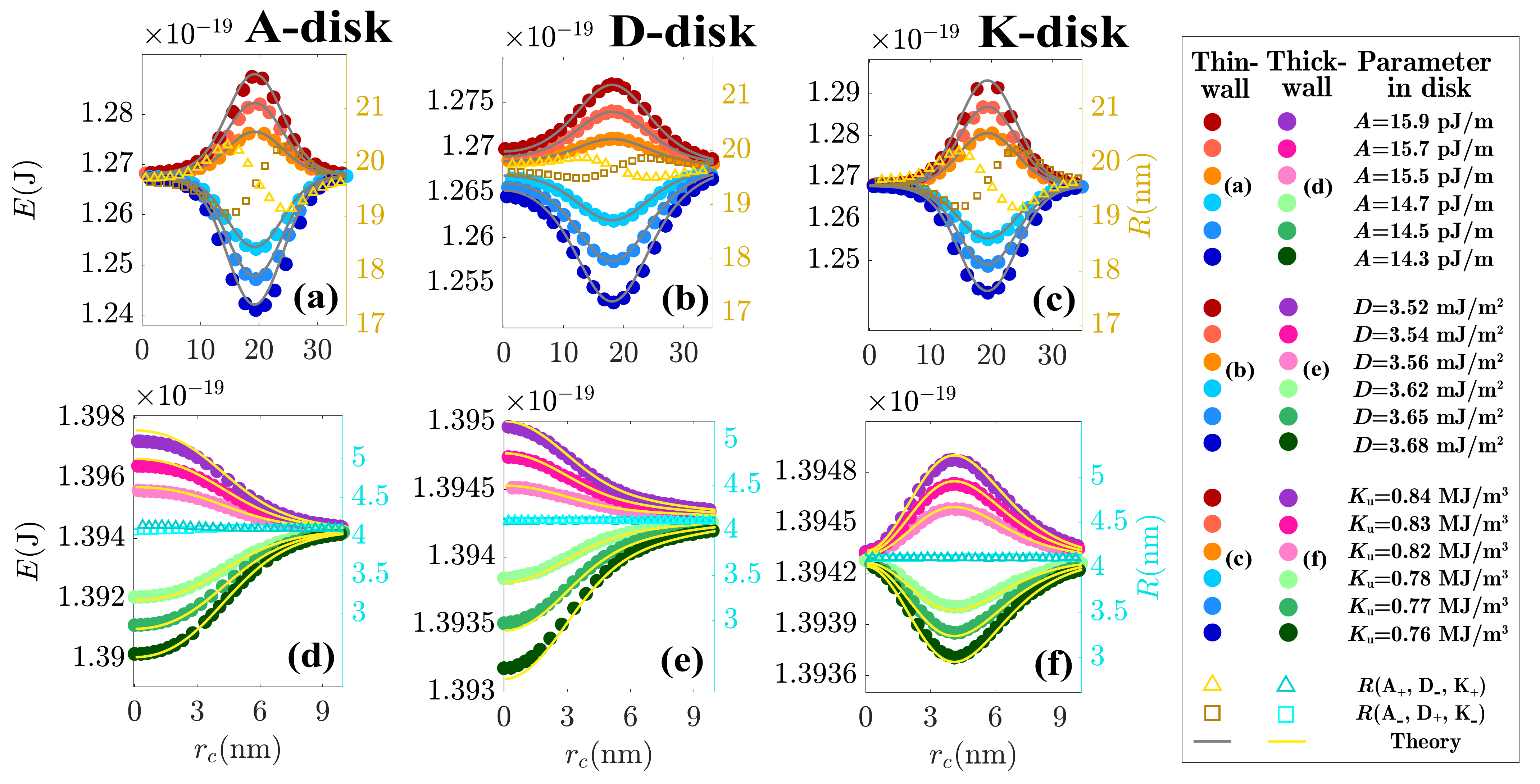}\\
\caption{Total film energy $E$ (solid dots) as a function of skyrmion center position 
$r_c$ for thin-wall skyrmions (a-c) and thick-wall skyrmions (d-f) in various small disks: 
A-disks (a,d), D-disks (b,e), and K-disks (c,f). Solid curves are the numerical 
integrations of Eq. (\ref{deltaE_small_disk}), with constant $R=19.57$ nm, $w=4.84$ nm 
for (a-c) and $R=4.10$ nm, $w=3.44$ nm for (d-f). The open squares and triangles 
are the skyrmion size as functions of $r_c$ in various disks.}
	\label{fig3}
\end{figure*}

In order to reveal pinning position differences for a thin-wall skyrmion and a 
thick-wall skyrmion in a small disk, we consider first skyrmion pinning 
by a $R_d=5 \ \mathrm{nm}$ disk, much smaller than a thin-wall skyrmion of 
$R= 19.57\ \mathrm{nm}$ with $\eta=4.04$. Initially, the thin-wall skyrmion is 
placed near the disk center with $r_c\equiv|\vec{r}_c|=0$, or far from the disk 
with $r_c\gg R$, or skyrmion rim is near the disk center with $r_c \approx R$. 
The skyrmion reaches its pinning position and stable structure through energy 
minimization, and the real pinning trajectories are obtained from 
running LLG equation.. Figures \ref{fig3} (a)-(c) show the total magnetic energy as 
functions of $r_c$. Symbols are the simulation results and curves are the theoretical 
value of Eq. (\ref{energy}) with a rigid skyrmion profile Eq. (\ref{profile}). 
The good agreement between the theoretical calculation and the Mumax3 
simulations shows that rigid approximation applies in this case.

For A$_-$, D$_+$ and K$_-$ disks, the energy minimum appears at $r_c\approx R$, 
as indicated by the potential wells of three lower curves in Figs. \ref{fig3} 
(a), (b) and (c), respectively. Hence disks with smaller exchange stiffness, 
larger DMI strength and weaker anisotropy pin skyrmions off-center. 
The results can be understood as follows. A skyrmion can lower its energy by placing 
as much as its wall regions inside A$_-$, D$_+$ and K$_-$ disks because any other 
alternative ways will increase the exchange energy of the skyrmion wall in an 
A$_-$ disk, or the DMI energy in a D$_+$ disk, or the anisotropy energy of a 
K$_-$ disk while the other types of energies are the same.
In contrast, energy is a local maximal when skyrmion is around $r_c=R$ and a local 
minimal around disk center ($r_c=0$) in A$_+$, D$_-$ and K$_+$ disks, see three 
top curves in Figs. \ref{fig3} (a), (b) and (c). Thus, a thin-wall skyrmion initially 
located at $r_c<R$ eventually stabilizes at the origin, or the disk center. 
However, if the skyrmion is initially at $r_c>R$, it will be expelled far away 
from the small disk because skyrmion energy there is the global minimal. 

The rigid skyrmion approximation is further checked by looking at the skyrmion size as a 
function of skyrmion center $r_c$ shown by the open squares in Figs. \ref{fig3} for 
A$_-$ ($A_1=14.7 \,\mathrm{pJ}/\mathrm{m}$) (a), D$_+$ ($D_1=3.62 \mathrm{\,mJ}/\mathrm
{m}^2 $) (b), and K$_-$ ($K_{\rm u,1}= 0.78  \,\mathrm{MJ}/\mathrm{m}^3$) (c) disk.
As the skyrmion center moves away from the disk center, the change of skyrmion size 
is no more than 2.5\% and the greatest skyrmion deformation occurs where $|\partial E
/\partial r_c|$ is maximal. Thus, the rigid approximation works well for small disks. 
The skyrmion radius decreases (increases) through a shrinkage (expansion) because a 
``mutual attraction" between small disks and skyrmion wall.
In addition, the skyrmion radii in other three type disks are plotted as the open 
triangles in Figs. \ref{fig3} (a)-(c), where a ``mutual repulsion" between 
small disks and skyrmion wall leads to the opposite behaviors.
Although the effect is negligible for small disks, it can be dominant for large disks,
which will be shown in the later sections.  

To investigate thick-wall pinning by a small disk, we use a perpendicular magnetic field 
to covert a thin-wall skyrmion to a thick-wall skrymion as explained in previous section. 
The skyrmion size becomes 4.1 nm under a field of $B=0.56$T with  $\eta=1.12$. 
Correspondingly, a disk of $R_d=1$ nm is used such that $R_d/R\gg1$ is kept the same.
The simulation procedure is the same as that for thin-wall skyrmions, and the results are 
plotted in Figs. \ref{fig3} (d)-(f). For A$_-$ and D$_+$ disks, the energy minimum moves 
to $r_c=0$, as illustrated in Figs. \ref{fig3} (d) and (e). Hence thick-wall skyrmions 
are pinned at the disk center or the symmetric point, in contrast to the off-center 
(asymmetric) pinning for thin-wall skyrmions. For A$_+$ and D$_-$ disks, the skyrmion energy 
at disk centers is maximal. Thick-wall skyrmions are then expelled away from disks, in 
contrast to the possible pinning at center for thin-wall skyrmions. Interestingly, K$_-$ 
(K$_+$) disks pin thick-wall skyrmions off-center (at center), similar to thin-wall ones. 
Also, the size of thick-wall skyrmions does not vary with $r_c$, due to the fact that 
thick-wall skyrmions have negligible cores and nearly constant wall width, thus making 
them incompressible. These results explain well why a magnetic field can drive skyrmion 
pinning position from off-center to center, and vice versa in many previous studies when 
A- and D-disks are used, but not for K-disks. 

\subsection{Physics of skyrmion pinning by small disks}

\begin{figure*}
	\centering
	\includegraphics[width=17.5cm]{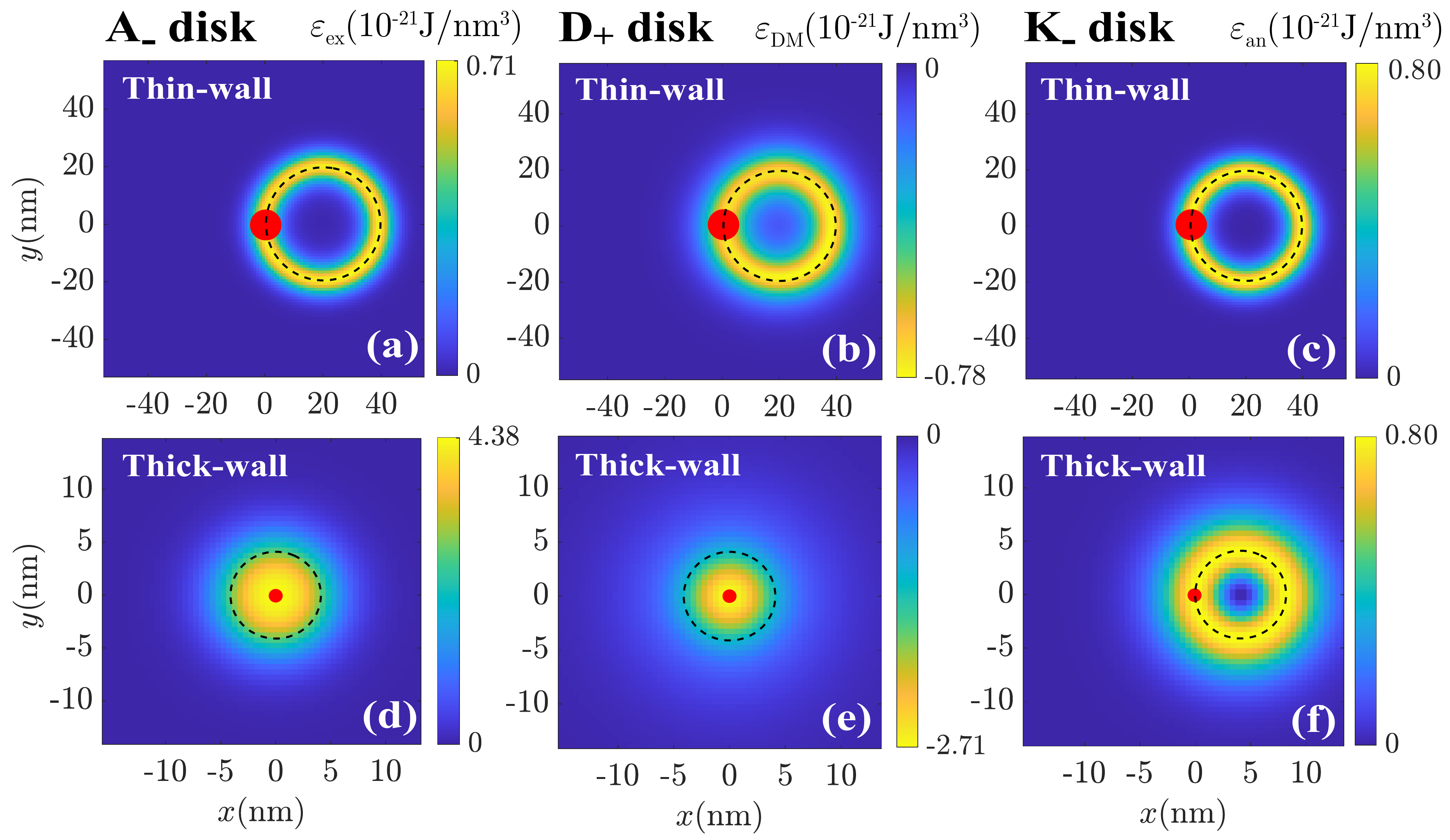}\\
\caption{Densities plots of exchange (a,d), DMI (b,e) and magnetic anisotropy (c,f) 
energies of a rigid skyrmion in a film. (a-c): thin-wall skyrmions, (d-f): thick-wall 
skyrmions. The color bars encode the the energy density (in the unit of $10^{-21}\ 
{\rm J/nm^3}$) and the black dashed circles denote the skyrmion wall center of $m_z=0$ 
that indicate the relative position to the disks denoted by the red solid circles: 
A$_-$ (a,d), D$_+$ (b,e), and K$_-$ (c,f) disks, respectively. 
Due to the symmetry, the pinning configurations are equivalent when the system is 
rotating around the skyrmion center.}
	\label{fig4}
\end{figure*}

The observed skyrmion pinning described in previous subsections can be understood from 
Lowest configurations. In the presence of a small foreign disk, the important quantity 
is total energy difference between the case of part of the rigid skyrmion on the disk 
and the case of a skyrmion in the film is 
\begin{equation}\label{deltaE_small_disk}
\Delta E  = d \iint_{\rm disk'}\left( \frac{\Delta A}{A} 
\varepsilon_{\rm ex} + \frac{\Delta D}{D} \varepsilon_{\rm DM} +  
\frac{\Delta K_{\rm u}}{K_{\rm u}} \varepsilon_{\rm an} \right) dxdy,
\end{equation}
where $\Delta A$, $\Delta D $, $\Delta K_{\rm u}$ are the difference of exchange 
stiffness, DMI strength and anisotropy constant between the disk and the film. 
In $\varepsilon_{\rm ex}$, $\varepsilon_{\rm DM}$ and $\varepsilon_{\rm an}$,
the rigid approximation gives rise to $A(x,y)\equiv A$, $D(x,y)\equiv D$ 
and $K_{\rm u}(x,y)\equiv K_{\rm u}$, respectively. The integration is over 
the part ($\text{disk}'$) of the disk where the ferromagnetic state is replaced 
by a part of the rigid skyrmion. By analyzing the gain and loss of total 
energy $\Delta E$, the pinning behaviors can be understood.

How $\varepsilon_{\rm ex}$, $\varepsilon_{\rm DM}$ and $\varepsilon_{\rm an}$ distribute 
over a film with a disk are plotted in Fig. \ref{fig4}. Subfigures (a)-(c) and (d)-(f) are 
for thin-wall and thick-wall skyrmions, respectively. In all plots, the yellow (blue) areas 
denote the regions where the absolute value of energy density is the highest (lowest), and the 
system tries to place the disk either in yellow parts or blue parts, depending on disk types.  
As an example, for a small A-disk, the total energy difference is about $\Delta E  = d 
(\Delta A/A) \iint_{\rm disk'} \varepsilon_{\rm ex}dxdy$. The low energy state is to place 
an A$_-$ (A$_+$) disk with $\Delta A<0$ ($\Delta A>0$) in the yellow (blue) regions. 
Hence thin-wall (thick-wall) skyrmions prefer to be pinned off-center (at-center)
of an A$_-$ disk, and at-center (faraway from the disk) of an A$_+$ disk.  
Similar analysis applies to D, K-disks. For D-disks, the low energy state is to place a 
D$_+$ (D$_-$) disk with $\Delta D>0$ ($\Delta D<0$) in the yellow (blue) regions, 
resulting in similar pinning preference as A-disks.  For K$_-$ (K$_+$) disks, 
the thin-wall and thick-wall skyrmions share similar  ``shining ring" energy density 
distribution, thus both of them are pinned off-center (at-center).

\begin{figure}
	\centering
	\includegraphics[width=8.5cm]{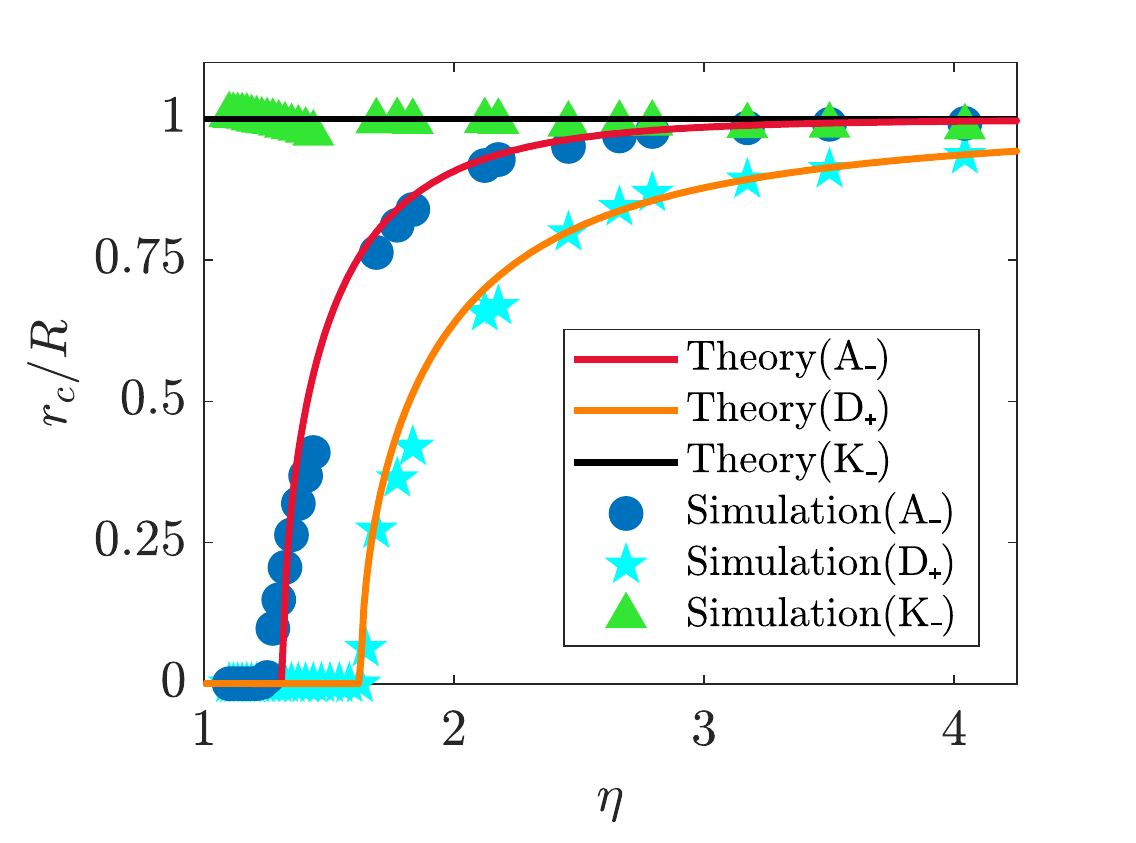}\\
	\caption{Skyrmion pinning positions as functions of $\eta$ for small A$_-$(D$_+$,K$_-$) disks.}
	\label{fig5}
\end{figure}

So far, we have focused on two extreme cases: thin-wall skyrmion
with $\eta\approx 4$ and thick-wall skyrmion with $\eta\approx 1$.
By varying $\vec{B}=B\hat{z}$, the skyrmion structure changes continuously.
For small A$_-$ disks, the dependence of pinning position on $\eta$ is shown 
by the blue dots in Fig. \ref{fig5}.
Interestingly, a critical $\eta$ around 1.31 exists.
Skyrmions with $\eta<1.31$ are pinned at disk center.
For those with $\eta>1.31$, their pinning position departs 
from $r_c=0$ and eventually takes $r_c=R$ for a large enough $\eta$.
Similarly, for small D$_+$ disks the critical $\eta$ is around 1.62 
(see the cyan stars in Fig. \ref{fig5}).
On the contrary, there is no shift in pinning position for small K$_-$ disks.
The off-center pinning of skyrmions always occurs at $r_c=R$ (the green 
triangles in Fig. \ref{fig5}). The simulation data coincide very well with
theoretical predictions from the energy minima positions (the solid curves 
in Fig. \ref{fig5}).

\subsection{Temperature dependence on skyrmion pinning lifetime}

\begin{figure}
\centering
\includegraphics[width=8.5cm]{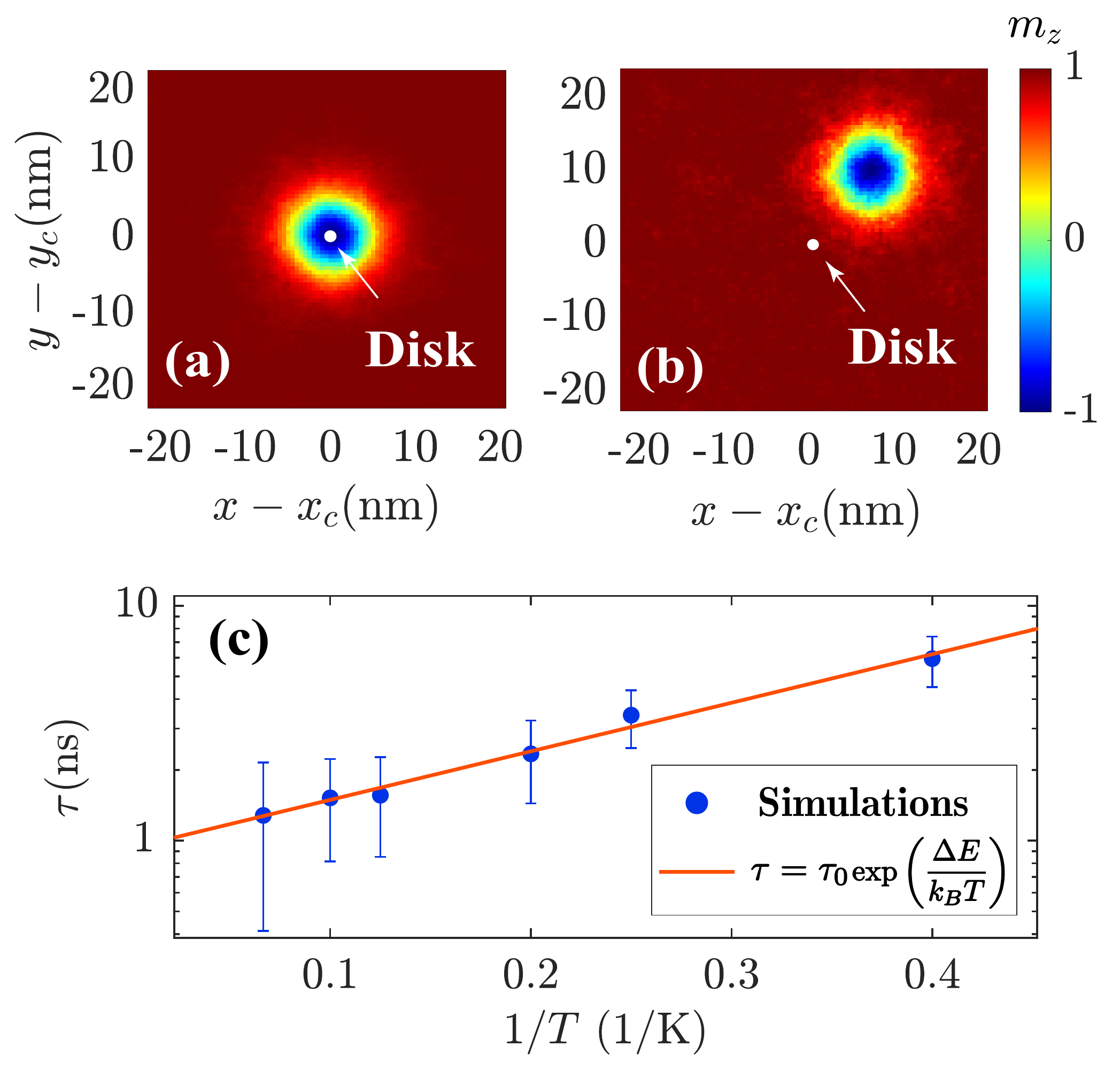}\\
\caption{Snapshots of a pinned skyrmion at $T=2.5 \mathrm{K}$ (a) and a 
depinned one at $T=15 \mathrm{K}$ (b) in a $\mathrm{K}_+$ disk of 
$K_{\mathrm{u}}=0.84 \mathrm{MJ}/\mathrm{m}^3$.
(c) The average lifetime $\tau$ of a pinned thick-wall skyrmion in a semi-logarithmic 
scale as a function of of $1/T$. The symbols are numerical data and the solid line is 
the Arrhenius form with $\tau_0=0.93 \,\mathrm{ns}$ and $\Delta E/k_B=4.76\mathrm{K}$.}
	\label{fig6}
\end{figure}

In the case that a skyrmion pinning site is a local energy minimal and the skyrmion 
far from the disk is the global energy minimal, the thermal agitation has strong 
effects on skyrmion pinning. Skyrmions can be depinned by the thermal fluctuations 
and diffuse to their global energy sites. To see how a skyrmion depins from a disk 
in this case, we take the thick-wall skyrmion pinned by a $K_{+}$ disk with 
$K_{\mathrm{u}}=0.84 \mathrm{MJ}/\mathrm{m}^3$ as an example. The other system 
parameters are the same as those for the top curve of Fig. \ref{fig3} (f). 
Referring to the potential landscape of the top curve shown in Fig. \ref{fig3} (f), 
a skyrmion is initially at center, which is the local energy minimal point with a 
potential barrier of $\Delta E=6\times 10^{-23}$J, corresponding to an activation 
temperature of $T = \Delta E/k_B = 4.35 \mathrm{K}$, here $k_B$ is the Boltzmann constant. 
Using MuMax3 to solve the stochastic LLG equation mentioned in the Model and Methods with 
temperature ranging from $2 \mathrm{K}$ to $20 \mathrm{K}$, we compute the average lifetime 
of the skyrmion pinned at the disk center. It is found that the skyrmion is pinned at the 
disk center at $T = 2.5 \mathrm{K}$ for a long time ($>$ 3 ns) but is quickly ($\sim$ 1 ns) depinned at 
$T = 15 \mathrm{K}$, snapshots of skyrmion and disk position are shown as in Figs. \ref{fig6} 
(a) and (b), indicating shorter skyrmion pinning lifetime at higher temperature.
The temperature dependent lifetime results are plotted in Fig. \ref{fig6} in 
which the $x-$axis is the inverse of the temperature in $\mathrm{K}^{-1}$ and $y-$axis 
is the semi-logarithmic of skyrmion lifetime $\tau$ in nano-second. 
Each data point is the average of 10 ensembles. 
The numerical data can be described well by the Arrhenius’s law
\begin{equation}
\tau (T) =\tau_{0}e^\frac{\Delta E}{k_B T}
\end{equation}
or 
\begin{equation}
\ln (\tau) =\frac{\Delta E}{k_B T} + \ln\tau_{0}
\label{Arrhenius}
\end{equation}
where $\Delta E$ is the activation energy and the inverse of prefactor $\tau_0^{-1}$ is 
the attempting frequency. The fitting of the numerical data to Eq. \eqref{Arrhenius} 
yield $\Delta E =6.6 \times 10^{-23} \mathrm{J}$, corresponding to activation temperature 
about $4.76 \, \mathrm{K}$ that is close to the theoretical value of $4.35 \, \mathrm{K}$ from the top 
potential landscape in Fig. \ref{fig3}(f). Numerically, we obtained $\tau_{0}=0.93 \, \mathrm{ns}$. 

In most skyrmion-based applications,  information bits are encoded by the existence and 
non-existence of skyrmions, and the operation speed is related to bit switching frequency.
Here, the attempting frequency $\tau_{0}^{-1}$ is about $1.1 \mathrm{GHz}$, and
the frequency $\tau^{-1}$  increases with a smaller potential barrier $\Delta E$. 
Thus, one may need a small $\Delta E$ for a high-speed operation on skyrmion.
These results should be helpful in the future design of skyrmionic devices.

\subsection{Gigantic skyrmion deformation}

\begin{figure*}
	\centering
	\includegraphics[width=18cm]{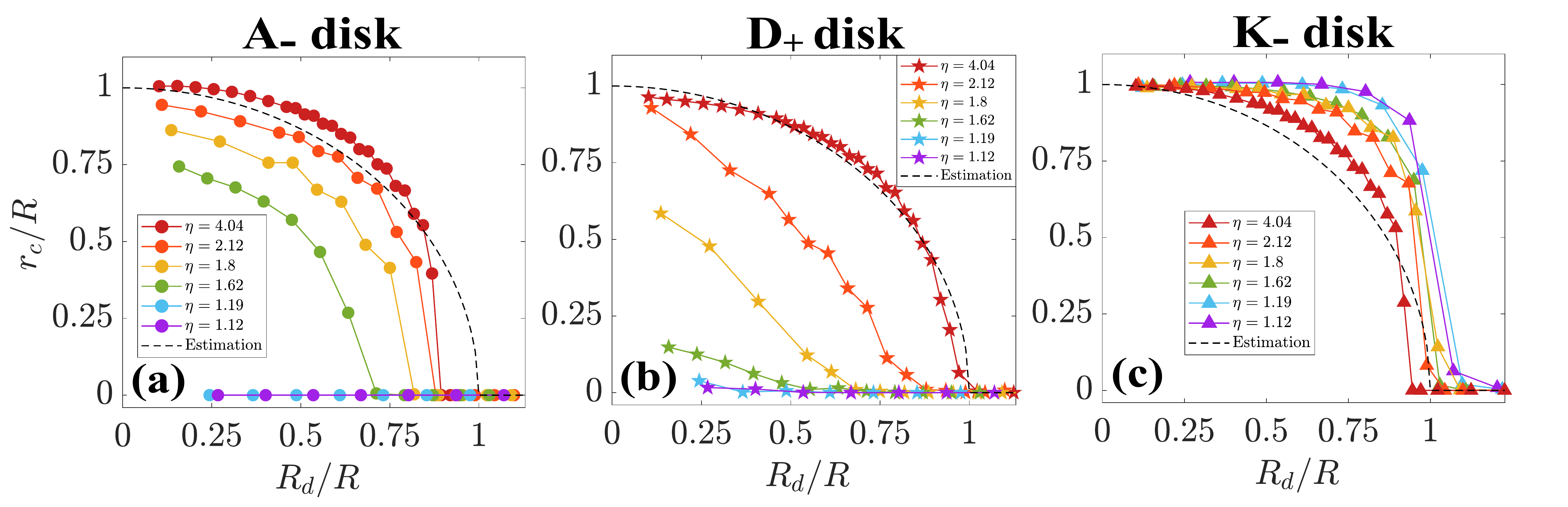}\\
	\caption{Pinning position $r_c$ (in units of $R$) as a function of $R_d$ 
	(in units of $R$) for an A$_-$ (a), a D$_+$ (b) and a K$_-$ (c) disks and for various $\eta$.  
The dashed lines are $r_c=\sqrt{R^2 - R_d^2}$.}
	\label{fig7}
\end{figure*}

\begin{figure*}
	\centering
	\includegraphics[width=18cm]{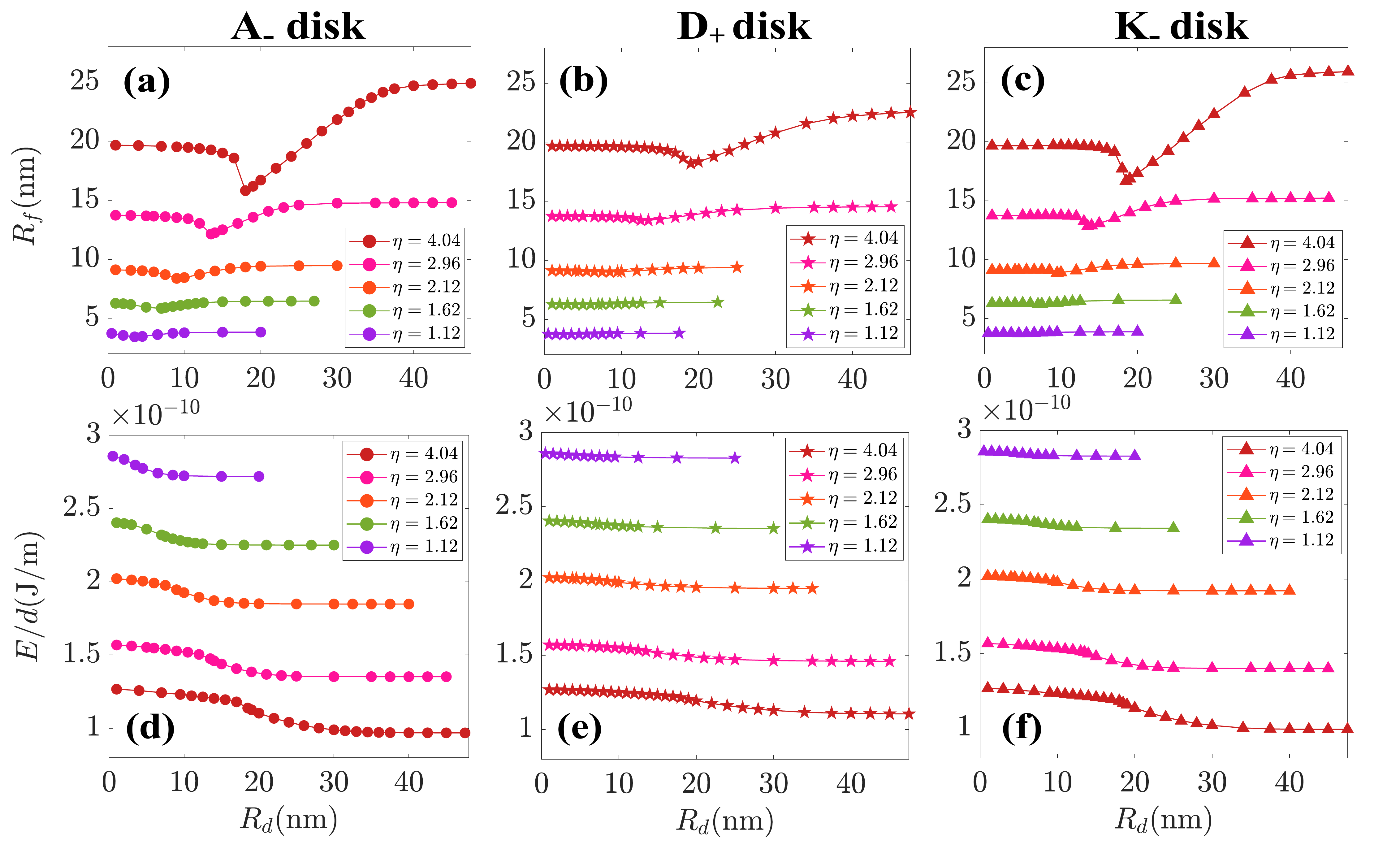}\\
\caption{The size $R_f$ (a-c) and total energy (d-f) of the deformed skyrmions of various $\eta$  
as functions of disk size $R_d$	in an A$_-$ (a,d), a D$_+$ (b,e), and a K$_-$ (c,f) disks .}
	\label{fig8}
\end{figure*}

Skyrmion behaves quite differently when a disk size is comparable to the size 
of the pinned skyrmions of large $\eta$. How the pinning position $r_c$ varies 
with an A$_-$ disk of size $R_d$ for various $\eta$, ranging from 1.12 to 4.04, 
is shown in Fig. \ref{fig7} (a), in units of original skyrmion radius $R$.
Deformed skyrmion size $R_f$ ($\ne R$ for large disks) is plotted in Fig. \ref{fig8} 
(a) as a function of $R_d$. The pinning of skyrmions with $\eta \ge 1.62$ behaves the 
same as a thin-wall skyrmion describe earlier. As the disk size increases, $r_c$ 
gradually changes from $R$ to 0 at a critical disk size slightly less than $R$. 
Near this critical disk size, the deformation of a thin-wall skyrmion is gigantic. 
The deformation increases with $\eta$ as shown in Fig. \ref{fig7}.  The thin-wall 
skyrmion shrinks itself such that the entire skyrmion is inside the disk in order 
to take the advantage of lower magnetic energy density of the disk. 
It appears that the disk ``swallows" the entire skyrmion. 
As the disk increases further, the skyrmion expands itself continuously to take the 
advantages of low energy land. This skyrmion deformation process occurs whenever the 
energy gain from sitting in the lower energy land is larger than the energy cost due 
to skyrmion deformation from its natural structure. Of course, when the process starts 
and ends depends on the $\Delta A$ and other material parameters. 

If the skyrmion pinning follows the rule of ``placing as much of the skyrmion wall as 
possible inside the disk", the skyrmion pinning position is given by $r_c=\sqrt{R^2-R_d^2}$, 
which is the dash line in Fig. \ref{fig7} (a), (b) and (c) for A$_-$, D$_+$ and K$_-$ disks.
It coincides with simulation data qualitatively before skyrmion deformation starts.
The agreement is better for skyrmions with larger $\eta$, showing the correctness of 
the rule of ``placing maximal skyrmion wall inside disks". 
Deformation of thick-wall skyrmions ($\eta \le 1.19$) by large disks is relatively weak.
As $R_d$ increases, the pinning is persistently at disk center, and no swallow of 
the skyrmion by a disk was observed as shown in Figs. \ref{fig7} and \ref{fig8}. 
In summary, the deformation of skyrmions can be understood from the view of energy competition.
On one hand, the skyrmion deforms itself such that it can fill the entire disk to 
take advantage of the lower energy density land. 
On the other hand, the skyrmion energy shall increase when it deviates from its natural shape. 
Their competition determines the shrinkage or expansion.

\subsection{Pinning and deformation in large A$_+$(D$_-$, K$_+$) disks}

\begin{figure}
\centering
\includegraphics[width=8.5cm]{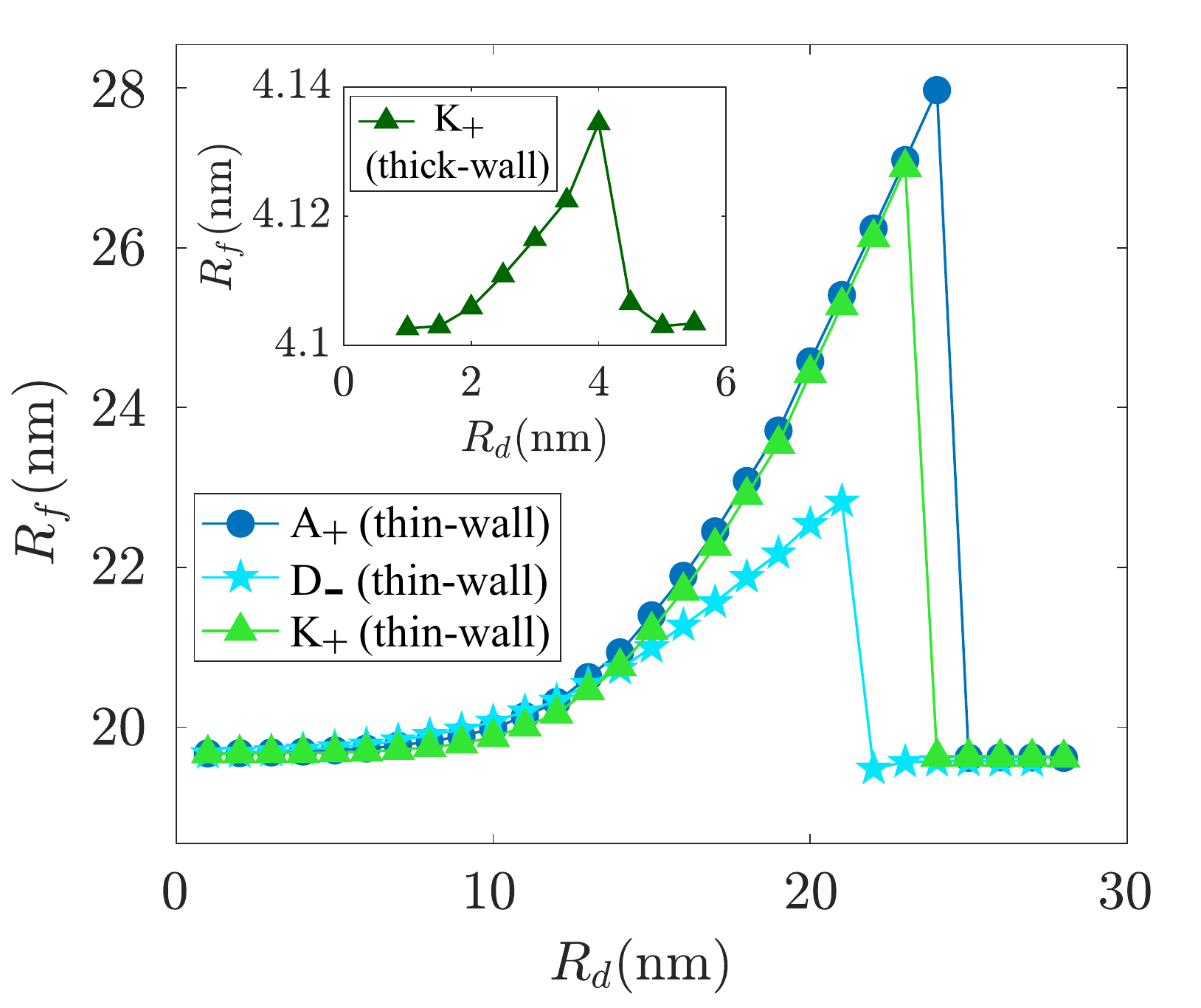}\\
\caption{Radius of a thin-wall skyrmion	as a function of disk size for A$_+$(D$_-$, K$_+$)	disks. 
	Inset: radius of a thick-wall skyrmion as a function of K$_+$ disk size.}
\label{fig9}
\end{figure}

Skyrmion pinning and deformation in large A$_+$(D$_-$, K$_+$) disks are quite different from those 
in A$_-$(D$_+$, K$_-$) disks. Let us use a K$_+$ disk as an example to explain the observations.
For thin-wall skyrmions, they are pinned at the centers of small K$_+$ disks with their 
walls staying outside of the disk. As the disk size increases, the skyrmion expands 
itself to keep its wall away from the high energy land of the disk. 
For the particular set of material parameters used in our simulations, the skyrmion 
keep expanding up to $R_d\approx 1.2 R$, as shown by the green triangles in Fig. \ref{fig9}.
A further increase of the disk size makes the energy cost of skyrmion expansion too high. 
Instead, it is more favourable to expell the skyrmion entirely out of the disk and return 
to its original structure and size. Similar behaviors were observed for large A$_+$ and 
D$_-$ disks as shown by the blue dots and the cyan stars in Fig. \ref{fig9}. 

\begin{figure*}
	\centering
	\includegraphics[width=18cm]{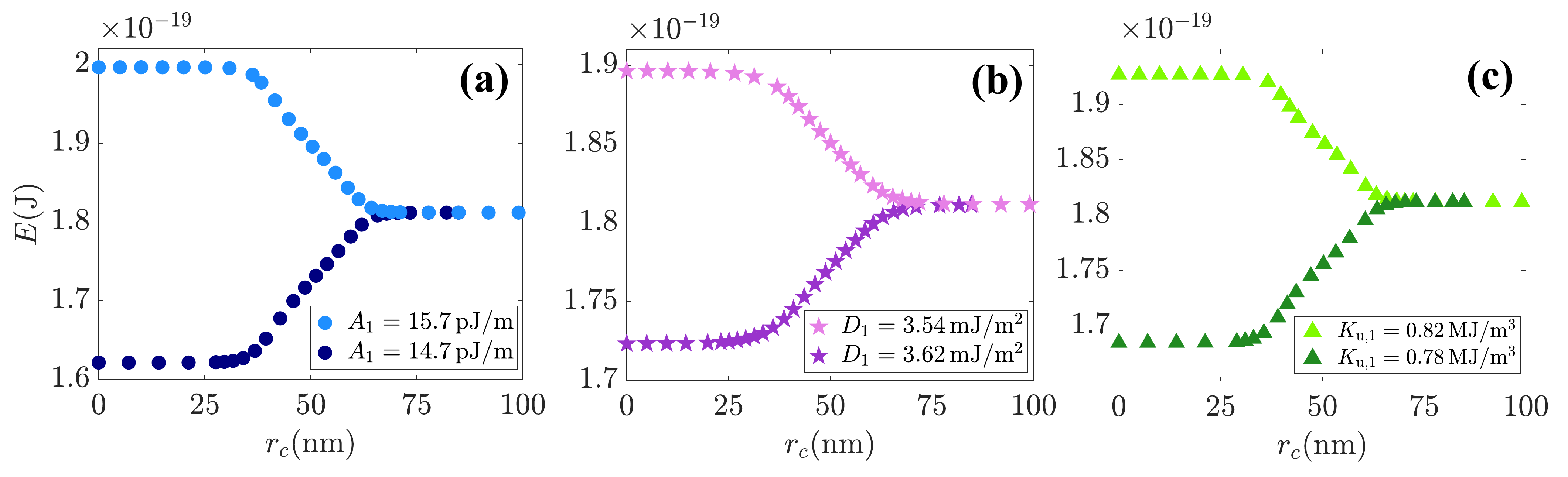}\\
\caption{Total energy of a skyrmion of $R=10.97$ nm and $w=4.46$ nm as a function of its 
center position in the presence of a large (a) A-, (b) D- and (c) K- disk of $R_d=50$ nm. 
The dark (light) colors are for A$_-$(D$_+$, K$_-$) [A$_+$(D$_-$, K$_+$)] disks. }
	\label{fig10}
\end{figure*}

As for thick-wall skyrmions, the expansion and expulsion are quite similar for a large 
K$_+$ disk as shown in the inset of Fig. \ref{fig9}. However, thick-wall skyrmions in 
A$_+$ and D$_-$ disks do not behave as their counterpart of thin-wall skyrmions. 
In fact, skyrmions cannot be pinned by A$_+$ and D$_-$ disks, and the thick-wall 
skyrmions are always expelled by those disks. The expulsion of skyrmions by A$_+$(D$_-$, 
K$_+$) disks can be understood by the energy dependence of $r_c$ for large disks.
The data are plotted in Fig. \ref{fig10}, with $R=10.97$ nm, $w=4.46$ nm and $R_d=50$ nm 
so that $R_d\gg R$. The dark-blue dots, the dark-purple stars and the dark-green triangles 
are skyrmion energies from simulation for A$_-$, D$_+$ and K$_-$ disks, respectively.
The light-colored ones are for A$_+$(D$_-$, K$_+$) disks. Clearly, for A$_-$(D$_+$, K$_-$) 
disks, the flat energy basins around $r_c=0$ imply pinning at the disk center. 
The flat energy plateaus around $r_c=0$ for A$_+$(D$_-$, K$_+$) disks means no pinning.

\section{Discussions and conclusion}
\label{sec4} 

Before conclusion, we would like to make several remarks. 
1) There is only one $r_c$ close to the disk at which the skyrmion energy is minimal 
for small A$_-$(D$_+$, K$_-$) disks while the skyrmion has a local minimal energy at 
a finite $r_c$ and global minimal energy at $r_c=\infty$ for small A$_+$(D$_-$, K$_+$) disks. 
Thus, in the case of A$_-$(D$_+$, K$_-$) disks, skyrmions will eventually pinned at 
the minimal energy site. In contrast, skyrmions can only be pinned by A$_+$(D$_-$, K$_+$) 
disks when they are near the local minimal region unless external stimuli can help the 
skyrmion to overcome the energy barrier. Otherwise, the skyrmion will be expelled far away 
from the disk. 
2) $A$ and $K_{\rm u}$ of realistic chiral magnets for skyrmions are always positive, but 
$D$ may be both positive and negative. The sign of $D$ will not change physics since it 
changes only chirality without change the energy and skyrmion size \cite{haitao2}. 
3) The disks used in this study are realizable. There are reports \cite{material1,material2,
material3,Hanneken2016} that exchange stiffness and anisotropy constant can be modified by 
doping. DMI constant can be modified by local coating or depositing other heavy metal or 
metal oxides. With today’s technologies, this can be done at least at sub-nanometer accuracy. 
4) The physics and results reported here are also applicable to other nano-structures with 
various shapes (linear, polygonal or even annular). Of course, the quantitive details may 
be different.  
5) When the pinned skyrmion size is comparable with the disk, the skyrmion will experience a 
gigantic structure deformation, which is crucial for intentional design of pinning structures 
with optimal size. These results should be important to skyrmion manipulation and in 
skyrmion-based spintronic application.

In summary, we have presented a unified picture of skyrmion pinning by a disk 
with $A$, $D$ or $K_{\rm u}$ differing from those of the embedded film. 
Depending on skyrmion structure, the type and relative size of the disk, 
a skyrmion can be pinned at the center or off-center of the disk. 
The physics is that skyrmion wall is more sensitive to the exchange stiffness 
and DMI constant although it can also sense the magnetic anisotropy. 
The skyrmion core mostly probes the magnetic anisotropy. Whether the total 
energy is lowered by putting the skyrmion wall or core inside a disk determines
the pinning at the center or off-center. It is shown that a thin-wall skyrmion 
is pinned off-center (at center) by a small A-disk with a weaker (stronger) 
exchange stiffness, or a small D-disk with a larger (smaller) DMI constant. 
On the other hand, a thick-wall skyrmion is pinned at the center of a small A-disk 
with a weaker exchange stiffness, or a small D-disk with a higher DMI constant, 
or a K-disk with stronger magnetic anisotropy. It can also be pinned off-center 
by a K-disk with weaker magnetic anisotropy. However, an A-disk with larger 
exchange stiffness or a D-disk with weaker DMI is not capable of pinning a skyrmion. 
We show that the rigid skyrmion approximation works well for small disks. 
A theory capable of accurately predicting the pinning position and skyrmion-disk 
interaction energy is presented. In the presence of temperature, skyrmion pinning by a local energy minimum has a finite lifetime that follows an Arrhenius law. 
When the disk size is comparable to that of a 
skyrmion, the skyrmion can lower the total energy by deforming itself such that 
the whole skyrmion is sucked into the disk. These results should be useful 
in fine tune of a skyrmion in skyrmion-based nanodevices and applications.

\section*{Acknowledgments}
X. Gong and K.Y. Jing contributed equally to this work. 
This work is supported by the National Natural Science Foundation of China 
(Grant No. 11974296) and Hong Kong RGC Grants (No. 16301518, 16301619 and 6302321).


\begin{thebibliography}{99}
	
	\bibitem{Muhlbauer2009}
	S. M\"{u}hlbauer, B. Binz, F. Jonietz, C. Pfleiderer, 
	A. Rosch, A. Neubauer, R. Georgii, and P. B\"{o}ni,
	Skyrmion lattice in a chiral magnet, 
	\href{https://science.sciencemag.org/content/323/5916/915.abstract}
	{Science {\bf{323}}, 915 (2009)}.
	
	\bibitem{YuXZ2010}
	X. Z. Yu, Y. Onose, N. Kanazawa, J. H. Park, J. H. Han, Y. Matsui, N. Nagaosa, and Y. Tokura,
	Real-space observation of a two-dimensional skyrmion crystal, 
	\href{https://www.nature.com/articles/nature09124}
     {Nature {\bf{465}}, 901 (2010)}.
	
	\bibitem{Bogdanov2006}
	U. K. R\"{o}\ss ler, A. N. Bogdanov, and C. Pfleiderer, 
	Spontaneous skyrmion ground states in magnetic metals, 
	\href{https://www.nature.com/articles/nature05056}
     {Nature (London) {\bf442}, 797 (2006)}.
	
	\bibitem{JiangW2015}
     W. Jiang, P. Upadhyaya, W. Zhang, G. Yu, M. B. Jungfleisch, F. Y. Fradin, J. E. Pearson, 
     Y. Tserkovnyak, K. L. Wang, O. Heinonen, S. G. E. te Velthuis, and A. Hoffmann,
	Blowing magnetic skyrmion bubbles, 
	\href{https://science.sciencemag.org/content/349/6245/283.abstract}
	{Science {\bf349}, 283 (2015)}.
	
	\bibitem{Sampaio2013}
	J. Sampaio, V. Cros, S. Rohart, A. Thiaville, and A. Fert, 
	Nucleation, stability and current-induced motion of isolated magnetic skyrmions in nanostructures,
	\href{https://www.nature.com/articles/nnano.2013.210}
	{Nat. Nanotechnol. {\bf{8}}, 839 (2013)}.
	
	\bibitem{Heinze2011}
	S. Heinze, K. von Bergmann, M. Menzel, J. Brede, A. Kubetzka, R. Wiesendanger, G. Bihlmayer, and S. Bl\"{u}ge,
	Spontaneous atomic-scale magnetic skyrmion lattice in two dimensions,
	\href{https://www.nature.com/articles/nphys2045}
	{Nat. Phys. {\bf 7}, 713 (2011)}.
		
	\bibitem{Boulle2016}
      O. Boulle, J. Vogel, H. Yang, S. Pizzini, D. S. Chaves, A. Locatelli, T. O. Mente\c{s}, A. Sala, 
	L. D. Buda-Prejbeanu, O. Klein, M. Belmeguenai, Y. Roussign\'{e}, A. Stashkevich, S. Mourad Ch\'{e}rif, 
	L. Aballe, M. Foerster, M. Chshiev, S. Auffret, I. M. Miron, and G. Gaudin, 
	Room-temperature chiral magnetic skyrmions in ultrathin magnetic nanostructures,
	\href{https://www.nature.com/articles/nnano.2015.315}{Nat. Nanotechnol. {\bf 11}, 449 (2016)}.
	
	\bibitem{Klaui2016}
	S. Woo, K. Litzius, B. Kr\"{u}ger, M.-Y. Im, L. Caretta, K. Richter, M. Mann, A. Krone, R. M. Reeve, 
	M. Weigand, P. Agrawal, I. Lemesh, M.-A. Mawass, P. Fischer, M. Kl\"{a}ui, and G. S. D. Beach,
	Observation of room-temperature magnetic skyrmions
	and their current-driven dynamics in ultrathin metallic ferromagnets, 
	\href{https://www.nature.com/articles/nmat4593}
	{Nat. Mater. {\bf 15}, 501 (2016)}.
	
	\bibitem{Parkin2020}
	A. K. Srivastava, P. Devi, A. K. Sharma, T. P. Ma, H. Deniz, H. L. Meyerheim, C. Felser, and S. S. P. Parkin,
	Observation of Robust N\'{e}el Skyrmions in Metallic PtMnGa,
	\href{https://doi.org/10.1002/adma.201904327}
	{Adv. Mater. {\bf 32}, 1904327 (2020)}.
	
	\bibitem{Fert2013}
	A. Fert, V. Cros, and J. Sampaio, Skyrmions on the track, 
	\href{https://www.nature.com/articles/nnano.2013.29}
	{Nat. Nanotechnol. {\bf8}, 152 (2013)}.

	\bibitem{Tomasello2014}
	R. Tomasello, E. Martinez, R. Zivieri, L. Torres, M. Carpentieri, and G. Finocchio,
	A strategy for the design of skyrmion racetrack memories,
	\href{https://www.nature.com/articles/srep06784}
	{Sci. Rep. {\bf 4}, 6784 (2014)}.
		
	\bibitem{Rohart2013}
	S. Rohart and A. Thiaville, Skyrmion confinement in ultrathin film nanostructures
	in the presence of Dzyaloshinskii-Moriya interaction, 
	\href{https://journals.aps.org/prb/abstract/10.1103/PhysRevB.88.184422}
	{Phys. Rev. B {\bf 88}, 184422 (2013)}.
	
	\bibitem{Nagaosa2013}
	N. Nagaosa and Y. Tokura, Topological properties and dynamics of magnetic skyrmions,
	\href{https://www.nature.com/articles/nnano.2013.243}
	{Nat. Nanotechnol. {\bf 8}, 899 (2013)}.
	
	\bibitem{ZhangXC2015}
	X. C. Zhang, Y. Zhou, M. Ezawa, G. P. Zhao, and W. S. Zhao, Magnetic skyrmion transistor: 
     skyrmion motion in a voltage-gated nanotrack,
	\href{https://www.nature.com/articles/srep11369}
	{Sci. Rep. {\bf 5}, 11369 (2015)}.
	
	\bibitem{Martinez2018}
	J. C. Martinez, W. S. Lew, W. L. Gan, and M. B. A. Jalil,
	Theory of current-induced skyrmion dynamics close to a boundary,
	\href{https://doi.org/10.1016/j.jmmm.2018.06.031}{J. Magn. Magn. Mater. {\bf 465}, 685 (2018)}.
		
	\bibitem{Ding2015}
	J. J. Ding, X. F. Yang, and T. Zhu,
	Manipulating current induced motion of magnetic skyrmions in the magnetic nanotrack,
	\href{https://iopscience.iop.org/article/10.1088/0022-3727/48/11/115004}
     {J. Phys. D: Appl. Phys. {\bf 48}, 115004 (2015)}.
		
	\bibitem{Queralt2019}
	J. Castell-Queralt, L. Gonz\'{a}lez-G\'{o}mez, N. Del-Valle,
	A. Sanchez, and C. Navau,
	Accelerating, guiding, and compressing skyrmions
	by defect rails,
	\href{https://pubs.rsc.org/en/content/articlelanding/2019/NR/C9NR02171J}
     {Nanoscale {\bf 11}, 12589 (2019)}.
	
	\bibitem{Iwasaki2013}
	J. Iwasaki, M. Mochizuki, and N. Nagaosa,
	Current-induced skyrmion dynamics in constricted geometries,
	\href{https://www.nature.com/articles/nnano.2013.176}
     {Nat. Nanotechnol. {\bf 8}, 742 (2013)}.
	
	\bibitem{Krause2016}
	S. Krause and R. Wiesendanger,
	Skyrmionics gets hot,
	\href{https://www.nature.com/articles/nmat4615}
	{Nat. Mater. {\bf 15}, 493 (2016)}.
	
	\bibitem{Kim2017}
	J.-V. Kim and M.-W. Yoo, Current-driven skyrmion dynamics in disordered films,
	\href{https://doi.org/10.1063/1.4979316}
	{Appl. Phys. Lett. 110, 132404 (2017)}.
		
	\bibitem{David2017}
	D. Cort\'{e}s-Ortu\~{n}o, W. W. Wang, M. Beg, R. A. Pepper, M. A. Bisotti, R. Carey, 
      M. Vousden, T. Kluyver, O. Hovorka, and H. Fangohr,
	Thermal stability and topological protection of skyrmions in nanotracks,
	\href{https://www.nature.com/articles/s41598-017-03391-8}
	{Sci. Rep. {\bf 7}, 4060 (2017)}.

	\bibitem{yuan1} H. Y. Yuan and X. R. Wang, 
	Domain wall pinning in notched nanowires, 
	\href{https://journals.aps.org/prb/abstract/10.1103/PhysRevB.89.054423}
	{Phys. Rev. B {\bf 89}, 054423 (2014)}.
%
%
	
	\bibitem{Hanneken2016}
	C. Hanneken, K. Kubetzka, A. von Bergmann, and R. Wiesendanger, 
	Pinning and movement of individual nanoscale magnetic skyrmions via defects,
	\href{https://iopscience.iop.org/article/10.1088/1367-2630/18/5/055009}
	{New J. Phys. {\bf{18}}, 055009 (2016)}.
	
	\bibitem{Lin2013}
	S. Z. Lin, C. Reichhardt, C. D. Batista, and A. Saxena, 
	Particle model for skyrmions in metallic chiral magnets: Dynamics, pinning, and creep,
	\href{https://journals.aps.org/prb/abstract/10.1103/PhysRevB.87.214419}
	{Phys. Rev. B {\bf{87}}, 214419 (2013)}.
	
	\bibitem{Muller2015}
	J. M\"{u}ller and A. Rosch, Capturing of a magnetic skyrmion with a hole,
	\href{https://journals.aps.org/prb/abstract/10.1103/PhysRevB.91.054410}
	{ Phys. Rev. B {\bf{91}}, 054410 (2015)}.
		
	\bibitem{Liu2013}
	Y. H. Liu and Y. Q. Li, A mechanism to pin skyrmions in chiral magnets,
	\href{https://iopscience.iop.org/article/10.1088/0953-8984/25/7/076005}
	{J. Phys. Condens. Matter. {\bf{25}}, 076005 (2013)}.
	
	\bibitem{Toscano2019}
	D. Toscano, S. A. Leonel, P. Z. Coura, and F. Sato, 
	Building traps for skyrmions by the incorporation of magnetic defects into nanomagnets: 
	Pinning and scattering traps by magnetic properties engineering,
	\href{https://www.sciencedirect.com/science/article/pii/S0304885318335686}
	{J. Magn. Magn. Mater. {\bf{480}}, 171 (2019)}.
	
	\bibitem{Song2021}
	C. Song, C. Jin, H. Xia, Y. Ma,
	J. Wang, Q. Liu, and J. Wang,
	Pinning and rotation of a skyrmion in Co nanodisk with nanoengineered point and ring defects, 
	\href{https://doi.org/10.1088/1361-648X/abda7e}
	{J. Phys. Condens. Matter. 117138 (2021)}.
		
	\bibitem{Dusan2017}
	D. Stosic, T. B. Ludermir, and M. V. Milo\v{s}evi\'{c}, 
	Pinning of magnetic skyrmions in a monolayer Co film on Pt(111): 
	Theoretical characterization and exemplified utilization,
	\href{https://link.aps.org/doi/10.1103/PhysRevB.96.214403}
	{Phys. Rev. B {\bf{96}}, 214403 (2017)}.
		
	\bibitem{Suess2020}
	D. Suess, C. Vogler, F. Bruckner, P. Heistracher, F. Slanovc, and C. Abert, 
	Spin Torque Efficiency and Analytic Error Rate Estimates of Skyrmion Racetrack Memory, 
	\href{https://www.nature.com/articles/s41598-019-41062-y}
	{Sci. Rep. {\bf 9}, 4827 (2019)}.

	\bibitem{Pathak2021}
	S. A. Pathak and R. Hertel, 
	Geometrically Constrained Skyrmions, 
	\href{https://doi.org/10.3390/magnetochemistry7020026}{Magnetochemistry {\bf 7}, 26 (2021)}.

	\bibitem{depin3} I. L. Fernandes, J. Chico, and S. Lounis, 
	Impurity-dependent gyrotropic motion, 
	deflection and pinning of current-driven ultrasmall skyrmions in PdFe/Ir(111) surface, 
	\href{https://iopscience.iop.org/article/10.1088/1361-648X/ab9cf0}
	{J. Phys. Condens. Matter. {\bf32}, 425802 (2020)}.
	
	\bibitem{Fernandes2019}
	I. L. Fernandes, J. Bouaziz, S. Bl\"{u}gel, and S. Lounis, 
	Universality of defect-skyrmion interaction profiles, 
	\href{https://www.nature.com/articles/s41467-018-06827-5}
	{Nat. Commun. {\bf 9}, 4395 (2018)}
	
	\bibitem{Arjana2020}
	I. G. Arjana, I. L. Fernandes, J. Chico, and S. Lounis, 
	Sub-nanoscale atom-by-atom crafting of skyrmion-defect interaction profiles, 
	\href{https://www.nature.com/articles/s41598-020-71232-2}
	{Sci. Rep. {\bf 10}, 14655 (2020)}.
	
	\bibitem{GongX2020}
	X. Gong, H. Y. Yuan, and X. R. Wang, 
	Current-driven skyrmion motion in granular films,
	\href{https://link.aps.org/doi/10.1103/PhysRevB.101.064421}
	{Phys. Rev. B {\bf{101}}, 064421 (2020)}.
	
	\bibitem{JingKY2021}
	K. Y. Jing, C. Wang, and X. R. Wang, 
	Random walk of antiferromagnetic skyrmions in granular films, 
	\href{https://journals.aps.org/prb/abstract/10.1103/PhysRevB.103.174430}
	{Phys. Rev. B {\bf103}, 174430 (2021)}.	
	
	\bibitem{CR2015}
	C. Reichhardt, D. Ray, and C. J. O. Reichhardt,
	Collective Transport Properties of Driven Skyrmions with Random Disorder,
	\href{https://journals.aps.org/prl/abstract/10.1103/PhysRevLett.114.217202}
	{Phys. Rev. Lett. {\bf 114}, 217202 (2015)}.	
	
	\bibitem{skyrmionsize}
	X. S. Wang, H. Y. Yuan, and X. R. Wang, 
	A theory on skyrmion size,
	\href{https://www.nature.com/articles/s42005-018-0029-0}
	{Commun. Phys. {\bf {1}}, 31 (2018)}.
	
	\bibitem{mumax3}
	A. Vansteenkiste, J. Leliaert, M. Dvornik, M. Helsen, F. G. Sanchez, 
	and B. V. Waeyenberge, The design and verification of MuMax3,  
	\href{https://aip.scitation.org/doi/10.1063/1.4899186}
	{AIP Advances {\bf 4}, 107133 (2014)}.

	\bibitem{haitao2}
	H. T. Wu, X. C. Hu, K. Y. Jing and X. R. Wang, 
	Size and profile of skyrmions in skyrmion crystals, 
	\href{https://www.nature.com/articles/s42005-021-00716-y}{Commun. Phys. {\bf 4}, 210 (2021)}.
				
	\bibitem{Brown}W. F. Brown, 
	Thermal fluctuations of a single-domain particle. 
	\href{https://link.aps.org/doi/10.1103/PhysRev.130.1677}{Phys. Rev.  {\bf 130}, 1677 (1963)}.
 
	
%
	\bibitem{xrw2} X. R. Wang, P. Yan, and J. Lu,
	High-field domain wall propagation velocity in magnetic 
	nanowires,
	\href{https://iopscience.iop.org/article/10.1209/0295-5075/86/67001}
	{{Euro. Phys. Lett.} {\bf 86}, 67001 (2009)}.
	
	
	
	
	\bibitem{ptco1}
     P. J. Metaxas, J. P. Jamet, A. Mougin, M. Cormier, J. Ferré, V. Baltz, B. Rodmacq, B. Dieny, and R. L. Stamps,
	Creep and Flow Regimes of Magnetic Domain-Wall Motion in Ultrathin 
	Pt/Co/Pt Films with Perpendicular Anisotropy, 
	\href{https://journals.aps.org/prl/abstract/10.1103/PhysRevLett.99.217208}
	{Phys. Rev. Lett. {\bf 99}, 217208 (2007)}.
			
	
	
			\bibitem{material1}
	 S. A. Bunyaev, B. Budinska, R. Sachser, Q. Wang, K. Levchenko, S. Knauer,
	  A. V. Bondarenko, M. Urbánek, K. Y. Guslienko,  A. V. Chumak,  M. Huth, G. N. Kakazei,
	   and O. V. Dobrovolskiy, Engineered magnetization and exchange stiffness in direct-write Co–Fe nanoelements,
	\href{https://doi.org/10.1063/5.0036361}{Appl. Phys. Lett. {\bf 118}, 022408 (2021)}.
	
		\bibitem{material2}
	Jeroen Mulkers, Bartel Van Waeyenberge, and Milorad V. Milo\v{s}evi\'{c}, 
	Effects of spatially engineered Dzyaloshinskii-Moriya interaction in ferromagnetic films, 
	\href{https://journals.aps.org/prb/abstract/10.1103/PhysRevB.95.144401}{Phys. Rev. B {\bf 95}, 144401 (2017)}.
	
\bibitem{material3}
	A. L. Balk, K. Kim, D. T. Pierce, M. D. Stiles, J. Unguris, and S. M. Stavis,
	Simultaneous control of the Dzyaloshinskii-Moriya interaction and magnetic anisotropy in nanomagnetic trilayers, 
	\href{https://journals.aps.org/prl/abstract/10.1103/PhysRevLett.119.077205}{Phys. Rev. Lett. {\bf 119}, 077205 (2017)}
	

	
	
	
	
	
\end{thebibliography}
\end{document}